\documentclass[twocolumn,showpacs,preprintnumbers,amsmath,amssymb]{revtex4}

\usepackage{graphicx} % Include figure files
\usepackage{dcolumn} % Align table columns on decimal point
\usepackage{bm} % bold math

\begin{document}

\preprint{}

\title{Multiple Andreev Reflections in a Carbon Nanotube Quantum Dot}

\author{M.\ R.\ Buitelaar}
\altaffiliation[Present address : ]{Cavendish Laboratory,
Cambridge, UK}
\author{W.\ Belzig}
\author{T.\ Nussbaumer}
\author{B.\ Babi\'c}
\author{C.\ Bruder}
\author{C. Sch{\"o}nenberger}
\email{Christian.Schoenenberger@unibas.ch} %\homepage{www.unibas.ch/phys-meso}
\affiliation{Institut f\"ur Physik, Universit\"at Basel, Klingelbergstrasse~82,
CH-4056 Basel, Switzerland }
\date{\today}% It is always \today, today,

\begin{abstract}
We report resonant multiple Andreev reflections in a multiwall
carbon nanotube quantum dot coupled to superconducting leads. The
position and magnitude of the subharmonic gap structure is found
to depend strongly on the level positions of the single-electron
states which are adjusted with a gate electrode. We discuss a
theoretical model of the device and compare the calculated
differential conductance with the experimental data.
(pdf including figures, see: www.unibas.ch/phys-meso/Research/Papers/2003/MAR-MWNT.pdf)
\end{abstract}

% 74.78.Na Nanoscale superconducting systems
% 74.45.+c  Proximity effects; Andreev reflection
% 73.63.Kv Quantum dots
% 73.21.La Quantum dots
% 73.23.Hk Coulomb blockade; single-electron tunneling
% 73.63.Fg Nanotubes

\maketitle

% ------------------------------------------------------------------
% Main text
% -------------------------------------------------------------------

%%%%%%%%%
% Intro %
%%%%%%%%%

The electronic transport properties of quantum dots coupled to
metallic leads have been the object of extensive theoretical and
experimental study \cite{Kouwenhoven1}. When weakly coupled to its
leads, the low-temperature transport characteristics
%CS of a quantum dot
are dominated by size and charge quantization effects,
parameterized by the single-electron level spacing $\Delta E$ and
the single-electron charging energy $U_C$. When the coupling of
the quantum dot to the source and drain electrodes is increased,
higher-order tunneling processes such as the Kondo effect become
important \cite{Kouwenhoven2}. New effects are expected when the
leads coupled to the quantum dot are superconductors. In that case
electron transport at small bias voltages is mediated by multiple
Andreev reflection (MAR) \cite{Andreev,Blonder}. Unlike
conventional S-N-S devices, however, the MAR structure is now
expected \cite{Yeyati,Johansson1} to strongly depend on the level
positions of the single-electron states of the quantum dot which
can be tuned with a gate electrode. The influence of Coulomb and
Kondo correlations have been addressed theoretically in
Refs.~\cite{Zaikin}.

Because MAR is suppressed rapidly for low-transparency junctions,
its observation requires a relatively strong coupling between the
leads and quantum dot. Even more so as on-site Coulomb repulsion,
which is common to weakly coupled dots, is expected to reduce
Andreev processes even further. A quantum dot very weakly coupled
to superconducting leads has been studied experimentally by Ralph
\textit{et al.} \cite{Ralph}. In this case the transport
characteristics were indeed dominated by charging effects and MAR
was completely suppressed.

The coupling to the leads, expressed in the life-time broadening
$\Gamma$ of the levels of the quantum dot, should be compared to
the superconducting gap energy $\Delta$. Favourable for the
observation of MAR in a quantum dot are coupling strengths of
order $\Gamma \sim \Delta$ and a small charging energy $U_C <
\Delta$. Together with the restriction that $\Gamma < \Delta E$
(for any quantum dot) this leads to the approximate condition $U_C
\lesssim \Gamma \lesssim \Delta E$. For most
%CS (semiconductor)
quantum dots typically the opposite is true and $\Delta E \ll
U_C$. It has recently been shown \cite{Buitelaar1} however, that
well-coupled multiwall carbon nanotube (MWNT) quantum dots can
have favourable ratio's of $\Delta E / U_C \sim 2$ and $U_C$ can
be as small as 0.4 meV, comparable to the energy gap $2 \Delta$ of
a conventional superconductor like Al.

%%%%%%%%%%%%%%%%%%%%%%%%%%%%%%%%%%%%%%%%%%%
% Description normal state experimentally %
%%%%%%%%%%%%%%%%%%%%%%%%%%%%%%%%%%%%%%%%%%%

Here we report on the experimental study of resonant MAR in a MWNT
quantum dot. The superconducting leads to the MWNT consist of an
Au/Al bilayer (45/135 nm). Before investigating the system in the
superconducting state, the sample is first characterized in the
normal state by applying a small magnetic field. From these
measurements relevant parameters such as $\Delta E$, $U_C$, and
$\Gamma$ are obtained. We then discuss a theoretical model that
describes the differential conductance of an individual level in a
quantum dot coupled to superconducting electrodes. In the final
part of the paper we compare the calculated differential
conductance with the experimental data.

\begin{figure}
\caption{\label{Fig1}
\textbf{(a)} Greyscale representation of the differential
conductance as a function of source-drain ($V_{sd}$) and gate
voltage ($V_g$) at \mbox{$T= 50$\,mK} and \mbox{$B=26$\,mT} for a
MWNT quantum dot. Here and in the following greyscale plots, the
darker a region the higher the differentical conductance.  The
dashed lines outline the onset of first-order tunneling processes.
\textbf{(b)} Linear-response conductance $G$ as a function of
$V_g$. The appearance of a single broad peak is due to the Kondo
effect. \textbf{(c)} Differential conductance at two different
values of $V_g$.} \end{figure}

Figure \ref{Fig1} shows a greyscale representation of the
differential conductance $dI/dV_{sd}$ versus source-drain
($V_{sd}$) and gate voltage ($V_g$) at \mbox{$T = 50$\,mK} when
the contacts are driven normal by a small magnetic field. The
dotted white lines outline the onset of first-order tunneling and
appear when a discrete energy level of the quantum dot is at
resonance with the electrochemical potential of one of the leads.
>From these and other electron states measured for this sample
(about $20$ in total), we obtain an average single-electron level
spacing \mbox{$\Delta E \sim 0.6$\,meV} and a charging energy
\mbox{$U_C \sim 0.4$\,meV}. Since $U_C = e^2/C_\Sigma$ this yields
\mbox{$C_\Sigma = 400$\,aF} for the total capacitance which is the
sum of the gate capacitance ($C_g$) and the contact capacitances
$C_s$ (source) and $C_d$ (drain). From the data of
Fig.~\ref{Fig1}a we obtain $C_g/C_\Sigma = 0.0036$, and $C_s/C_d =
0.45$. The lifetime broadening $\Gamma$ is obtained from the width
of the single-electron peaks at finite source-drain voltage
(taking the background conductance into account) and is found to
be \mbox{$\Gamma \sim 0.35$\,meV}.

The high-conductance ridge around \mbox{$V_{sd}=0$\,mV} in
Fig.~\ref{Fig1}a is a manifestation of the spin-$1/2$ Kondo effect
occurring when the number of electrons on the dot is odd. As a
result, the Coulomb valley in the conductance has disappeared in
this region of $V_g$ and at \mbox{$50$\,mK}, which is the base
temperature of our dilution refrigerator, only a single peak
remains, see Fig.~\ref{Fig1}b. This is known as the unitary limit
of conductance. The appearance of the Kondo effect is an
indication that the coupling to the leads is relatively strong. We
will not discuss the Kondo effect here, and instead refer to
Ref.~\cite{Buitelaar2}.

%%%%%%%%%%%%%%%%%%%%%%%%
% Theory weak coupling %
%%%%%%%%%%%%%%%%%%%%%%%%

When the magnetic field is switched off the leads become
superconducting. To calculate the expected differential
conductance in the superconducting state of the leads we have used
the non-equilibrium Green-function technique \cite{Greenfunction}.
We model the quantum dot as a series of spin-degenerate resonant
levels coupled to superconducting electrodes, which are assumed to
have a BCS spectral density. Note, that neither electron-electron
interaction (Coulomb blockade) nor exchange correlations (Kondo
effect) are accounted for in the model, which may, therefore, not
explain all details of the actual measurements. However, the
interplay between MAR and resonant scattering already leads to
strongly nonlinear IV-characteristics and reproduces some of the
key features of the data.  The details of the model calculation
will be presented elsewhere \cite{Bruderbelzig}. The main
parameters entering the calculation are the two tunneling rates
$\Gamma_{s(d)}= 2\pi|t_{s(d)}|^2 N_0$, related to microscopic
tunneling amplitudes $t_{s(d)}$ via the density of states $N_0$ of
the source (drain) contact. In the model we account for the gate
voltage by a shift of the level, which can be adjusted according
to the experimentally observed Coulomb diamonds, see
Fig.~\ref{Fig1}.

\begin{figure}
\caption{\label{Fig2}
\textbf{(a)} Greyscale representation of the calculated
differential conductance as a function of $V_{sd}$ and level
position ($\varphi_g\propto V_g$) for a single-electron level
coupled symmetrically to superconducting electrodes. For clarity,
the low-energy part $e V_{sd} \lesssim 2 \Delta /4$ has been
omitted. The dashed lines indicate resonance positions, as
explained in the text. \textbf{(b)} Differential conductance at
$\varphi_g = 0$. Note, that MAR peaks at $2 \Delta /n$ are
suppressed for even values of $n$. \textbf{(c)} Schematics of a
single electron state between superconducting source and drain
electrodes. The situation shown corresponds to point $C$ in panel
(a). The spectral density is shown at the bottom. The Lorentzian
level broadening in the normal state is replaced by a narrow
central resonance accompanied by a series of satellite
peaks.\textbf{(d)} Same as (c) for point $D$ in panel (a).}
\end{figure}

The discrete nature of the single-electron states is most
pronounced when $\Gamma$ is small. Therefore, before presenting
the model calculation that directly compares to the experimental
data, we first discuss the transport characteristics of a single
spin-degenerate level with a relatively weak and symmetric
coupling to the superconducting leads along the lines of
Refs.~\cite{Yeyati,Johansson1}. The total tunneling rate $\Gamma
\equiv \Gamma_s+\Gamma_d$ is set to $\Delta$. Figure \ref{Fig2}
shows the corresponding greyscale representation of the calculated
differential conductance $dI/dV_{sd}$ versus $\varphi_g:=eV_g
C_g/C_{\Sigma}$ and $eV_{sd}$. The peak structure in $dI/dV_{sd}$
at $V_{sd} < 2 \Delta/e$ is the result of MAR. In general, Andreev
channels become available for transport at voltages $V_{sd} = 2
\Delta /ne$, where $n$ is an integer number. These positions are
indicated by the horizontal dashed black lines in
Fig.~\ref{Fig2}a. The appearance and magnitude of the MAR peaks,
however, is strongly dependent on the position of the resonant
level in the quantum dot with respect to the Fermi energy of the
leads. Only those MAR trajectories that connect the resonant level
to the leads' BCS spectral densities give a significant
contribution to the current. Consider, for example, the position
marked by $C$ in Fig.~\ref{Fig2}a, which corresponds to the
schematics of Fig.~\ref{Fig2}c and indicates the position
$(\varphi_g,eV_{sd})=(0,2 \Delta /3)$. The corresponding
second-order Andreev trajectory connects the gap edges of the
source and drain electrodes and includes the resonant level which
is situated exactly in between the respective Fermi energies. This
results in the large peak in $dI/dV_{sd}$ seen in
Fig.~\ref{Fig2}b.

A similar peak is absent at $(\varphi_g,eV_{sd})=(0,\Delta)$,
corresponding to point $D$ in Fig.~\ref{Fig2}a. At this voltage,
first-order Andreev reflection becomes possible. The corresponding
trajectories (see Fig.~\ref{Fig2}d), however, do not directly
connect the resonant level to the leads spectral densities, and
therefore do not significantly contribute to the current. Only
when the gate voltage is adjusted to align the level with the
Fermi energy of one of the leads (indicated by the arrows in
Fig.~\ref{Fig2}a) a peak in $dI/dV_{sd}$ is observed. It can be
shown (for symmetric junctions) that the subharmonic gap structure
at $V_g = 0$ is suppressed for all voltages $V_{sd} = 2 \Delta
/ne$ with $n = \textrm{even}$ \cite{Yeyati,Johansson1}. When
$V_{sd}$ is increased beyond $\Delta/e$, peaks are observed either
when the level stays aligned with the electrochemical potential of
the leads (red dashed lines: $\varphi_g =\pm eV_{sd}/2$) or when
the level follows the gap edges as an initial or final state of an
Andreev process (blue dash-dotted lines:
$\varphi_g=\pm(\Delta-eV_{sd}/2)$ or
$\varphi_g=\pm(\Delta-3eV_{sd}/2)$).

%%%%%%%%%%%%%%%%%%%%%%%%%%%%%%%%%%%%
% Data + Theory realistic coupling %
%%%%%%%%%%%%%%%%%%%%%%%%%%%%%%%%%%%%

\begin{figure}
\caption{\label{Fig3} \textbf{(a)}
Greyscale representation of the measured differential conductance
as a function of $V_{sd}$ and $V_g$ at \mbox{$T= 50$\,mK} with the
leads in the superconducting state. The gate voltage range
corresponds to the left part of Fig.~1a. The dashed white lines
emphasize the position of the MAR peaks. \textbf{(b-c)}
Differential conductance at the positions given in panel (a).}
\end{figure}

We now turn to the actual measurements of the differential
conductance when the leads are in the superconducting state.
Figure~\ref{Fig3} shows a greyscale representation of the measured
$dI/dV_{sd}$ versus $V_{sd}$ and $V_g$ at \mbox{$B=0$\,mT} for the
same single-electron state of Fig.~\ref{Fig1}. A number of
differences between the normal state (Fig.~\ref{Fig1}) and
superconducting state (Fig.~\ref{Fig3}) can be observed. The
horizontal high-conductance lines at \mbox{$V_{sd} = \pm 0.2$\,mV}
in Fig.~\ref{Fig3}, for example, are attributed to the onset of
quasi-particle tunneling when
% the voltage difference between the source and drain electrode equals
$V_{sd} = 2 \Delta /e$.
The subgap
structure at $V_{sd} < 2 \Delta /e$ (i.e. below \mbox{$0.2$\,mV})
is attributed to MAR. As anticipated, the magnitude and the
position (dashed white lines) of MAR peaks depend on $V_g$. To
allow for comparison with theory, the adjustable parameters of the
model are set to the values obtained from the measurement of
Fig.~\ref{Fig1}. The most important parameter is the coupling
between the electrodes and the dot which turns out to be
\mbox{$\Gamma \sim 3.5$\, $\Delta$}. The voltage division between
the two tunnel barriers separating the quantum dot from the leads
is $C_s / C_d = 0.45$. The individual tunneling rates
% of the source and drain electrodes to the dot
are not exactly known but are not expected to show a strong
asymmetry \cite{Kondo}. The
neighboring single-electron states, separated by $\Delta E \sim
6.5 \Delta$, are included in the calculation. The finite
temperature of the experiment is taken into account and set to
\mbox{$T=0.1$\,$\Delta$}.

\begin{figure}
\caption{\label{Fig4} \textbf{(a)}
Greyscale representation of the calculated differential
conductance as a function of $V_{sd}$ and level position
($\varphi_g$). The different adjustable parameters represent the
experimental situation, see text. \textbf{(b-c)} Differential
conductance at the positions given in panel (a). For the two
dashed curves, $\Gamma$ has been chosen approximately twice as
large as the experimental value.}
\end{figure}

The resulting calculated greyscale representation of the
differential conductance is shown in Fig.~\ref{Fig4}. The overall
appearance clearly resembles the measured data of Fig.~\ref{Fig3}.
For example, both the model and the measured data show a large
peak in $dI/dV_{sd}$ around \mbox{$V_{sd} = 0$\,mV} when the
electron state is at resonance with the electrochemical potential
of the leads (i.e. at $V_g = 0$). When the level is moved away
from this position, the linear-response conductance rapidly decays
to values below its normal-state value. In contrast, the
differential conductance peak at $V_{sd} = 2 \Delta/e$ shows the
opposite behavior (both in the model as in the experiment). At
$V_g = 0$, this peak is much less pronounced then at lower values
of $V_g$. This can also be seen in the $dI/dV_{sd}$ line traces of
Fig.~\ref{Fig3}b-c and Fig.~\ref{Fig4}b-c (solid curves). These
observations are similar to conventional S-N-S structures, such as
atomic-sized break junctions \cite{Scheer1}: for large
transparencies of the junction a peak is observed at $V_{sd} = 0$
but no structure at $2 \Delta$ while for small transparencies a
gap is observed around $V_{sd} = 0$ and a large peak at $2 \Delta$
marks the onset of quasi-particle tunneling. For these systems,
the transparency depends on the atomic arrangement of the
junction, here the effective transparency can be tuned by moving
the level position of a single-electron state with a gate
electrode. The effective transparency is large if the level is
aligned with the Fermi levels of the leads (on resonance) and it
is small otherwise (off resonance).

The subharmonic gap structure is clearly visible in the measured
data of Fig.~\ref{Fig3} and has a similar gate-voltage dependence
as in the model calculation of Fig.~\ref{Fig4}.  However, there
are several differences. The most dramatic one is the pronounced
peak at $(V_g,V_{sd})=(0,\Delta/e)$ in the measurement
(Fig.~\ref{Fig3}c). Because this position corresponds to an even
MAR cycle it should be absent based on our previous consideration
(see Fig.~\ref{Fig2}d).

Let us compare theory and experiment by focussing onto the
$dI/dV_{sd}$ lines traces shown in Fig.~\ref{Fig3}b-c and
Fig.~\ref{Fig4}b-c (solid curves). In panels b) the dot level is
off resonance, while it is on resonance in panels c). For the
former case, experiment and theory agree fairly well. The
differences in MAR structure between model and experiment are much
more pronounced at the resonance position. Whereas the experiment
(Fig.~\ref{Fig3}c) reveals pronounced peaks at $\Delta/2$,
$\Delta$ and $2\Delta$, the calculated $dI/dV_{sd}$
(Fig.~\ref{Fig4}c, solid) reveals fine structure for small
$V_{sd}$ and pronounced peaks at $2\Delta/3$ and $2\Delta$.
According to our previous discussion $dI/dV_{sd}$ should indeed
show a pronounced peak at $V_{sd}=2\Delta/3e$, if the dot level is
centered in the middle, i.e. for $\varphi_g=0$ at point C in
Fig.~\ref{Fig2}a and Fig.~\ref{Fig2}c. It rather appears in the
experiment that, contrary to expectations, the subgap feature at
$2 \Delta /3$ is missing, while the `forbidden' at $\Delta$ is
present. Such behavior would only be expected for very asymmetric
junctions having $C_s /C_d \ll 1$ \cite{Johansson2}, which is not
the case in the present work.

There are different imaginable scenarios that may account for the
observed $\Delta$ peak and the lack of fine structure around
\mbox{$V_{sd} = 0$\,mV} in the data of Fig.~\ref{Fig3}c. Inelastic
scattering processes inside the dot, for example, would broaden
and obscure higher-order MAR features. Other possible reasons may
be found in a broadened BCS spectral density (the superconductor
consists of a bilayer of Au/Al \cite{Scheer2}) or a suppression of
higher-order MAR due to the on-site Coulomb repulsion.

In a phenomenological approach, we may try to account for the
additional broadening by manually introducing larger bare
couplings $\Gamma_{s,d}$. Many curves with varying parameters were
calculated of which a representative set is displayed in
Fig.~\ref{Fig4}b-c (dashed curves) corresponding to relatively
large dot-electrode couplings of \mbox{$\Gamma_s=2.5$\,$\Delta$}
and \mbox{$\Gamma_d=3.5$\,$\Delta$}. For the off-resonance
position (Fig.~4b) the main effect of the larger $\Gamma$ is the
increased magnitude of $dI/dV_{sd}$. In contrast, the MAR
structures significantly changes for the resonance position
(Fig.~4c). Remarkably, at large coupling $\Gamma$, peaks now
appear at $2\Delta$, $\Delta$ and $\Delta/2$. These peaks do not
originate from the resonant level, but from the two neighboring
ones which are off resonance (the dot levels are spaced by
\mbox{$\Delta E \approx 6.5$\,$\Delta$}). Though the agreement is
now reasonable, there is one remaining problem. We were unable to
reproduce the relative peak height between the $2\Delta$ and
$\Delta$ peaks. Using any reasonable set of parameters, the
$2\Delta$ peaks is always larger than the $\Delta$ peak in the
model, while it is the opposite in the experiment. We emphasize
that the model does not take into account interaction and
correlations. Since a Kondo resonance is observed in the normal
state, which need not be suppressed in the superconducting
state~\cite{Buitelaar2}, this may be the origin of the
discrepancy. The Kondo resonance changes the spectral density in
the leads by adding spectral weight to the center of the gap and
removing spectral weight from the gap edges. The former tends to
enhance the $\Delta$ peak, while the latter tends to suppress the
$2\Delta$ one. This explanation is attractive, but more work both
in theory and experiment is needed to substantiate it.

%%%%%%%%%%%%%%%
% Conclusions %
%%%%%%%%%%%%%%%

In conclusion, we have investigated the non-linear conductance
characteristics of a quantum dot coupled to superconducting
electrodes. We find a strong dependence of the MAR structure on
the level position of the single-electron states. The experimental
data is compared with a theoretical model, assuming a BCS
density-of-states in the electrodes and an interaction-free dot.
Reasonable agreement is possible, if the tunneling coupling to the
leads is enhanced by a factor $\sim 2$ in the model as compared to
the experimental value. There are additional subtle differences
which point to the importance of interaction and exchange
correlations.

\begin{acknowledgments}
We acknowledge contributions by H. Scharf and M. Iqbal. We thank
L. Forr\'o for the MWNT material and J. Gobrecht for providing the
oxidized Si substrates. This work has been supported by the Swiss
NFS, BBW, and the NCCR on Nanoscience.
\end{acknowledgments}

% End of Main text
%---------------------------------------------------------------------------

%-------------------------------------------------------------

%-------------------------------------------------------------

\begin{thebibliography}{99}
\bibitem{Kondo}
The saturation of the Kondo resonance of Fig.~1 close to
\mbox{$2$\,$e^2/h$} suggests that the tunneling rates of both
junctions are quite similar.
\end{thebibliography}
\end{document}